\documentclass[aps,prl,reprint,superscriptaddress,nofootinbib,longbibliography,floatfix]{revtex4-2}
\usepackage{amsmath,amssymb,bm,mathtools,graphicx,microtype,mathrsfs,amsthm}
\usepackage{needspace}
\usepackage[hidelinks]{hyperref}
\newcommand{\TV}{\operatorname{TV}}
\newcommand{\Tr}{\operatorname{Tr}}
\newcommand{\E}{\mathbb E}
\newcommand{\id}{\mathbb I}

\newcommand{\ket}[1]{|#1\rangle}
\newcommand{\bra}[1]{\langle#1|}
\newcommand{\prlsection}[1]{\par\textit{#1.}---\,}
\newtheorem{theorem}{Theorem}
\makeatletter
\AtBeginDocument{\immediate\write\@auxout{\string\citation{apsrev42Control}}}
\makeatother
\begin{document}
\title{Classical Simulation of Lossy Gaussian Boson Sampling beyond Square-Root Regime}
\author{Youngrong Lim}
\email{sshaep@gmail.com}
\affiliation{Department of Physics, Chungbuk National University, Cheongju 28644, Korea}
\begin{abstract}
Photon loss makes Gaussian boson sampling easier to simulate classically. A standard approach replaces the lossy input by classical light, and it stays accurate while the mean number of surviving photons grows no faster than the square root of the number $N$ of squeezed inputs at fixed squeezing and accuracy. We prove that this scale is a limit of replacement by classical light. For explicit interferometers, no state with a nonnegative Glauber--Sudarshan $P$ representation, Gaussian or not and however correlated, keeps the error in the photon count distribution small beyond this scale. We then surpass this limit by approximating the detection instead of the input. The sampler evolves the state exactly between detections and, after each detection, replaces the conditional state by a Gaussian with the same covariance. For every interferometer its total variation error is $O(N\eta^3)$ at polynomial expected cost, where $\eta$ is the photon survival probability. The mean surviving photon number can therefore grow as $N^{2/3}$ at fixed accuracy. A separate pair construction reaches $N^{3/4}$ when every input mode is equally squeezed. 
\end{abstract}
\maketitle

\prlsection{Introduction}
Gaussian boson sampling (GBS) probes interference among photons using squeezed light, passive optics, and photon-number-resolving (PNR) detection~\cite{Aaronson2013,Lund2014,Hamilton2017,Kruse2019}, providing the basis for photonic quantum advantage experiments~\cite{Zhong2020,Zhong2021,Madsen2022,Deng2023,Liu2026Jiuzhang}. Photon loss is a dominant imperfection in experiments. It can also reduce the cost of classical simulation~\cite{Oszmaniec2018,GarciaPatron2019,Brod2020,Qi2020,Oh2021,Liu2023,Oh2024}. A claim of quantum advantage therefore has to check that the loss of the experiment lies outside the classically simulable regime. For Fock state inputs, efficient simulation is known when the surviving photon number grows more slowly than the square root of the input photon number~\cite{Oszmaniec2018,GarciaPatron2019}. A square-root simulation regime is also known for lossy GBS at fixed squeezing and accuracy~\cite{Qi2020,Liu2023,Oh2024}.

A nonnegative Glauber--Sudarshan $P$ representation gives an efficient PNR sampler whenever its distribution of coherent amplitudes can be sampled efficiently~\cite{Rahimi2016}. The classical Gaussian reference has a fidelity deficit $O(\eta^2)$ per input relative to the GBS state. For $N$ inputs, this yields a total variation bound $O(\sqrt N\,\eta)$~\cite{Qi2020,Fuchs1999}. At fixed squeezing and accuracy, this gives the sufficient regime $N\eta^2=O(1)$, with a suitably small constant, and hence $\overline N_{\rm out}=N\eta\sinh^2r=O(\sqrt N)$. However, Gaussian optimization alone does not establish whether non-Gaussian classical references can improve this exponent.

Tensor-network methods retain a nonclassical component. These methods decompose the lossy state exactly into a pure Gaussian state with reduced squeezing followed by correlated Gaussian displacement noise. The pure component is approximated by a matrix product state~\cite{Oh2024}. Under uniform loss its mean photon number is $N_{\rm eff}=\Theta(N\eta^2)$. The corresponding entanglement analysis retains the square-root leading power, with logarithmic refinements~\cite{Oh2024}. More recently, Liu \emph{et al.} developed a conversion from Gaussian states to matrix product states that avoids hafnian calculations~\cite{Liu2026}. 

We first show that allowing non-Gaussian classical references does not remove the square-root limitation. For explicit interferometers, every state with a nonnegative Glauber--Sudarshan $P$ representation incurs a total variation error of at least $1-e^{-c_rN\eta^2}$ in its PNR distribution, even with non-Gaussian references and arbitrary intermode correlations. Non-Gaussian classical references therefore cannot improve the square-root scaling in the worst case.

\begin{figure*}[t]
 \includegraphics[width=\textwidth]{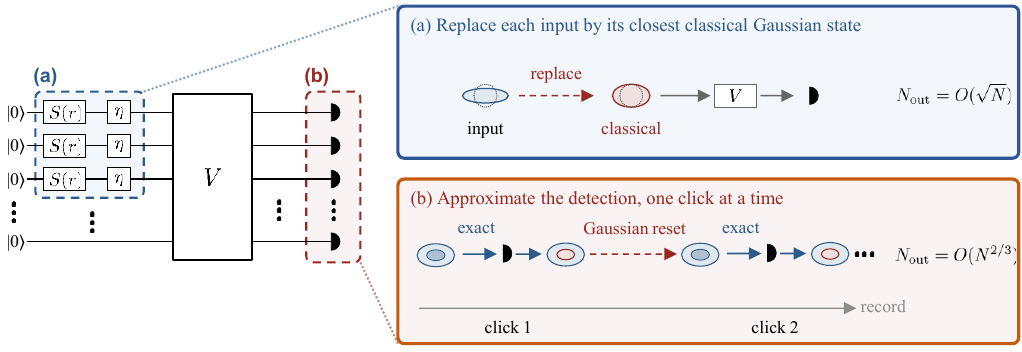}
 \caption{Classical-light replacement and detection approximation for lossy GBS. $N$ squeezed inputs $S(r)\ket{0}$ and $M-N$ vacua pass through loss $\eta$, an interferometer $V$, and PNR detectors. (a) In classical-light replacement~\cite{Qi2020} each lossy input, squeezed below the vacuum circle (dashed), is replaced by the classical Gaussian state of maximal fidelity and the rest is exact. Theorem~\ref{thm:lower} bounds the error from below by $1-e^{-c_rN\eta^2}$ on its interferometer family. (b) In the present method the input is exact and the detection record is processed one click at a time. After each click the non-Gaussian conditional state (red core) is replaced by the Gaussian with the same covariance at a cost of order $\eta^2$ for typical records, which gives Theorem~\ref{thm:cubic}.}
 \label{fig:schematic}
\end{figure*}

Our approach can surpass this bound by approximating the detection process rather than replacing the input by classical light (Fig.~\ref{fig:schematic}). The initial state is unchanged, and evolution between detections is treated exactly. At low transmission, a typical detection changes the remaining light only slightly. Retaining the leading change leaves an error of order $\eta^2$. This approximation is made only when a photon is detected, so loss supplies a second suppression. We prove a total variation error bound $O(N\eta^3)$ for the full PNR distribution, including rare records with large conditional fluctuations. The sampler has polynomial expected arithmetic cost. At fixed squeezing and accuracy, $N\eta^3=O(1)$ permits a mean surviving photon number of order $N^{2/3}$. When every input mode is equally squeezed, a simpler construction based on pair generation and amplification reaches $N^{3/4}$. The detection approximation also combines with the exact input decomposition of Ref.~\cite{Oh2024}. For the single-mode and few-mode examples studied, the combined sampler has lower empirical error than the classical Gaussian reference at parameter values motivated by experiments.

\prlsection{Setup}
The input consists of $N$ squeezed vacuum states with a common fixed magnitude $r>0$ and $M-N$ additional modes prepared in vacuum. The occupied columns of the passive interferometer form an isometry $V\in\mathbb C^{M\times N}$, $V^\dagger V=\id_N$. Squeezing phases are absorbed into its columns. Uniform pure loss of survival probability $\eta$ commutes with passive interference~\cite{Weedbrook2012}. Write
\begin{equation}
 n_0=\sinh^2r,\qquad \overline N_{\rm out}=N\eta n_0.
 \label{eq:model}
\end{equation}
The target $P_{\eta,V}$ is the full distribution on $\mathbb N^M$, and $\TV(P,Q)=\frac12\sum_{\bm n}|P(\bm n)-Q(\bm n)|$. The $M-N$ vacuum input modes are present before loss, and modes emptied by loss remain part of the $N$ squeezed inputs. Constants in the asymptotic bounds may depend on the fixed squeezing $r$.

\prlsection{Square-root barrier for classical light}
Let
\begin{equation}
 \mathcal C_M=\left\{\int\mu(d\bm\alpha)\,
 \ket{\bm\alpha}\bra{\bm\alpha}:\mu\ge0,\ \int\mu=1\right\}
 \label{eq:classical}
\end{equation}
be the $P$ classical states of $M$ modes~\cite{Yadin2018}, the mixtures of products of coherent states $\ket{\bm\alpha}=\bigotimes_i\ket{\alpha_i}$. The measure $\mu$ may be singular, non-Gaussian, and correlated across all modes, and it may depend on the target. Write $Q_\sigma$ for the PNR distribution of $\sigma\in\mathcal C_M$, the mixture under $\mu$ of products of Poisson distributions with intensities $|\alpha_i|^2$.

\begin{theorem}[Lower bound on $P$ classical replacement]
\label{thm:lower}
For every fixed $r>0$, even $N$, and $M\ge N$, there is a passive isometry $V_N$ such that
\begin{equation}
 \begin{gathered}
 \inf_{\sigma\in\mathcal C_M}
 \TV\!\left(P_{\eta,V_N},Q_\sigma\right)
 \ge1-e^{-c_rN\eta^2},\\
 c_r=\frac{\big[\sqrt{2n_0^2+n_0}-\sqrt2\,n_0\big]^2}{8(1+n_0)^3},
 \end{gathered}
 \label{eq:barrier}
\end{equation}
for $0\le\eta\le1$. On this family the optimal error within $\mathcal C_M$ is $\Theta_r(\min\{1,N\eta^2\})$.
\end{theorem}

\emph{Proof sketch.} $V_N$ sends inputs $2a-1$ and $2a$, $a=1,\dots,N/2$, through the beam splitter $\frac1{\sqrt2}\big(\begin{smallmatrix}1&i\\1&-i\end{smallmatrix}\big)$, which turns two equally squeezed vacua into a two-mode squeezed vacuum, and leaves the other $M-N$ outputs in vacuum. Let $p_{jk}$ be the probability that the two outputs of one pair record $j$ and $k$ photons under $P=P_{\eta,V_N}$, and $q_{jk}$ the same probability for a product coherent state with intensities $I$ and $J$. The counts of the coherent state are independent Poisson variables, so
\begin{equation}
 q_{11}-q_{20}-q_{02}=-\tfrac12e^{-I-J}(I-J)^2\le0,
 \label{eq:classicalwitness}
\end{equation}
a constraint on classical count probabilities~\cite{Lee1990,Caves1991}. Every $\sigma\in\mathcal C_M$ inherits it by mixing. The lossy pair violates it, as shown in the Supplemental Material (SM)~\cite{SM},
\begin{equation}
 p_{11}-p_{20}-p_{02}
 =\frac{n_0\eta^2}{[1+n_0(2\eta-\eta^2)]^2}>0,
 \label{eq:pairgap}
\end{equation}
at order $\eta^2$ and without any unbounded moment. To add up the pairs, let $h(1,1)=1$, $h(2,0)=h(0,2)=-1$, and $h=0$ otherwise, and let $W_t=\prod_a[1+th(n_{2a-1},n_{2a})]$ with $0<t<1$. At a fixed coherent amplitude of all $M$ modes the pair counts are independent, so Eq.~\eqref{eq:classicalwitness} gives $\E_{Q_\sigma}W_t\le1$ for every $\sigma\in\mathcal C_M$, whatever its correlations across pairs. Under $P$ the pairs are independent, and minimizing over $t$ gives $\E_PW_t^{-1}=(1-\Delta)^{N/2}$ with $\Delta=\tfrac12\big(\sqrt{p_{11}}-\sqrt{p_{20}+p_{02}}\big)^2\ge4c_r\eta^2$. The Cauchy--Schwarz inequality $\sum_{\bm n}\sqrt{PQ_\sigma}\le(\E_PW_t^{-1})^{1/2}(\E_{Q_\sigma}W_t)^{1/2}$ and $\TV\ge1-\sum_{\bm n}\sqrt{PQ_\sigma}$ then give
\begin{equation}
 \TV(P,Q_\sigma)\ge1-(1-\Delta)^{N/4}\ge1-e^{-c_rN\eta^2}.
 \label{eq:affinitylower}
\end{equation}
The product of thermal states with the marginals of $P$ is classical and gives the matching upper bound $N(n_0+3n_0^2)\eta^2$ derived in the SM.

Theorem~\ref{thm:lower} is a statement about a method, not about the hardness of a distribution. Every algorithm that replaces the input by classical light and then detects produces some $Q_\sigma$ with $\sigma\in\mathcal C_M$. On this family, every $Q_\sigma$ has a photon count error of at least $1-e^{-c_rN\eta^2}$. For fixed error below one this forces $N\eta^2=O_r(1)$, hence $\overline N_{\rm out}=O_r(\sqrt N)$. The square-root regime is therefore the best that classical-light replacement can reach in the worst case, whether the reference is Gaussian as in Ref.~\cite{Qi2020} or non-Gaussian and correlated across modes. The theorem does not cover references that are Gaussian but nonclassical, or the nonclassical core retained in Ref.~\cite{Oh2024}, and it is not a complexity lower bound.

\prlsection{Beyond the barrier by approximating the detection}
The pair family itself is easy to sample, as a product of two-mode distributions, and Theorem~\ref{thm:cubic} below samples it as a special case. The limitation is a property of the approximation class and not a lower bound on classical computation, so a method that keeps the input and approximates the detection is not bound by it.
\begin{theorem}[Classical sampling beyond the square-root barrier]
\label{thm:cubic} 
For every fixed $r>0$, $M\ge N$, and passive isometry $V$, there is a randomized classical algorithm with polynomial expected arithmetic cost whose output distribution $Q$ satisfies
\begin{equation}
 \TV(P_{\eta,V},Q)\le\min\{1,C_3(r)N\eta^3\},
 \qquad 0\le\eta\le1,
 \label{eq:mainthm}
\end{equation}
where $C_3(r)$ is independent of $M,N,V$. Thus $\eta=c_0N^{-1/3}$, for a sufficiently small fixed $c_0$, permits $\overline N_{\rm out}\propto N^{2/3}$ at fixed accuracy.
\end{theorem}
The theorem applies to the squeezed and vacuum inputs specified above, with no restriction on depth or connectivity, and output collisions are included.

\emph{Proof sketch.}
The construction rewrites PNR detection as a fictitious absorption process. Couple every output mode to a perfect absorber that removes photons one at a time and records the mode of each removal. Run the process until every photon is absorbed. The number of removals recorded in mode $i$ is then its photon number, so the record has exactly the PNR distribution of the state. The fictitious time only orders the clicks and is integrated out. Along the process the conditional state follows two rules. While no click occurs it is damped by $e^{-t\hat n/2}$, with $\hat n=\sum_ib_i^\dagger b_i$ the total photon number operator, which is the state conditioned on no photon having been removed yet. At a click in mode $i$ it becomes $b_i\rho b_i^\dagger/\lambda_i$, a photon subtraction, with $\lambda_i=\langle b_i^\dagger b_i\rangle$ the rate of that mode. Neither rule is a unitary operation, and nothing is applied in the experiment.

For a Gaussian state, the evolution between detections and the covariance change caused by a click have closed expressions. Use vacuum covariance $\id/2$, write $\Sigma=\id/2+K$, and set $m_0=\sinh r\cosh r$, $p_r=n_0+m_0$, and $d_r=m_0-n_0$. If $O$ is the realification of $V$, the state after loss and interference has
\begin{equation}
 K_0=\eta O\operatorname{diag}(p_r\id_N,-d_r\id_N)O^T.
 \label{eq:initial}
\end{equation}
The no-click damping keeps a Gaussian state Gaussian, and a photon subtraction changes its covariance by a known amount. For the projector $\Pi_i$ onto output $i$'s two quadratures,
\begin{equation}
 \begin{aligned}
 \dot K&=-K-K^2, &\lambda_i&=\tfrac12\Tr(\Pi_iK),\\
 K&\longmapsto K+\Delta_i, &\Delta_i&=\frac{K\Pi_iK}{\lambda_i}.
 \end{aligned}
 \label{eq:update}
\end{equation}
The waiting time to the next click also has a closed form. The probability of no click during the next interval of duration $t$ is
\begin{equation}
 S_K(t)=\det[\id+(1-e^{-t})K]^{-1/2},
 \label{eq:wait}
\end{equation}
and its limit $\det(\id+K)^{-1/2}$, the vacuum probability of the current state, is the probability that no photon is left.

The sampler, whose output distribution we denote by $Q_{\rm G}$, therefore stores only $K$. Starting from $K_0$, it draws from Eq.~\eqref{eq:wait} whether the record has ended and otherwise the time of the next click. It chooses the clicked output in proportion to the rates $\lambda_i$ and applies Eq.~\eqref{eq:update}. The clicks per output form the sample. The no-click step is exact. The single approximation is that after a click the exact state, which is no longer Gaussian, is replaced by the centered Gaussian whose covariance is $\id/2+K+\Delta_i$. Gaussian trajectories and photon subtraction formulas are established tools~\cite{Verstraelen2018,Walschaers2017}. We bound the accumulated error in the full PNR distribution for every interferometer, including rare detection records. No hafnian is needed, photon number and covariance are not capped, and the expected number of clicks is at most $N\eta n_0$. With $K=OXO^T$ stored through a $2N\times2N$ matrix $X$, a direct implementation has expected cost $O(MN+(1+N\eta n_0)(N^3+MN^2))$ and memory $O(MN+N^2)$, with scalar inversion of $S_K$ counted as a primitive, as derived in the SM.

The click rate and the error of one reset suggest the cubic scale. At low transmission each output holds $O(\eta)$ photons. A subtraction therefore changes the state by $\Delta_i=O(\eta)$, and along typical records the replacement by a Gaussian costs $O(\eta^2)$ per click. Clicks occur at a total rate of order $N\eta$, and the product is $N\eta^3$. The following bound also includes rare records with large covariance changes.

Let $\mathsf P=OO^T$, with rank $2N$, and define a shift used only in the proof
\begin{equation}
 b=\frac\eta{1-\eta},\qquad
 Y_t=K_t+be^{-t}\mathsf P,\qquad
 \Phi_t=\Tr(Y_tK_t^2)\ge0.
 \label{eq:potential}
\end{equation}
No noise is added to the sampler, and an exact vacuum direction contributes zero to $\Phi_t$. An estimate of the relative jump size gives, along every candidate trajectory,
\begin{equation}
 Y_t\succeq0,\qquad 0\preceq\Delta_i\preceq2Y_t,
 \qquad \sum_i\lambda_i\Delta_i=K_t^2.
 \label{eq:relative}
\end{equation}
These inequalities use the paired quadrature structure of squeezed inputs, not a lower bound on any output's occupation.

Let $e_i$ be the trace distance between the exact state after a click and its Gaussian replacement. The local estimate based on Gaussian displacements implies
\begin{equation}
 \begin{gathered}
 \sum_i\lambda_i e_i\le A_{r,\eta}\Phi_t,\qquad
 A_{r,\eta}=\frac{2C_*}{(1/2-\eta d_r)^2},\\
 C_*<0.268069.
 \end{gathered}
 \label{eq:localcost}
\end{equation}
Let $\mathscr G$ be the generator of the process, so that $\mathscr G\Phi$ is the expected rate of change of $\Phi_t$. It is the derivative along the no-click flow, including the time derivative of the shift, plus the jump terms $\sum_i\lambda_i[\Phi(K+\Delta_i)-\Phi(K)]$. It obeys
\begin{equation}
 \mathscr G\Phi\le -3\Phi+10\Tr(Y^2K^2)
 \le(-3+10\|Y\|)\Phi.
 \label{eq:drift}
\end{equation}
The linear jump terms cancel the nonlinear drift between detections by Eq.~\eqref{eq:relative}, and noncommuting jumps are included.

Choose $b<a<3/10$ and let $\mathcal T$ be the first time $\|Y_t\|\ge a$. Before that time the drift is negative, and at a crossing $\Phi_{\mathcal T}\ge a(a-b)^2$. Consequently,
\begin{equation}
 a(a-b)^2\Pr(\mathcal T<\infty)
 +(3-10a)\E\int_0^{\mathcal T}\Phi_t\,dt\le\Phi_0.
 \label{eq:budget}
\end{equation}
Compare the target with a hybrid that uses the trajectory approximation up to $\mathcal T$ and exact readout afterward. Its error before the crossing is bounded by Eq.~\eqref{eq:localcost}, and its distance from the actual, unstopped sampler is at most the crossing probability. Thus
\begin{equation}
 \TV(P,Q_{\rm G})\le
 \min\!\left\{1,\Phi_0\max\!\left[
 \frac1{a(a-b)^2},\frac{A_{r,\eta}}{3-10a}\right]\right\},
 \label{eq:shortcert}
\end{equation}
with the explicit initial budget
\begin{equation}
 \Phi_0=N\eta^3\left[
 8n_0^3+6n_0^2+\frac{4n_0^2+2n_0}{1-\eta}\right].
 \label{eq:phi0}
\end{equation}
For $0<\eta\le1/10$, $a=8/27$ gives an explicit constant $C_{3,0}(r)$ in the SM. Taking $C_3(r)=\max\{C_{3,0}(r),10^3\}$ extends the bound to all $\eta$, trivially when $\eta>1/10$. The case $\eta=0$ is exact. The factor $N$ rather than $M$ comes from the occupied covariance subspace. Equation~\eqref{eq:budget} controls large conditional fluctuations with the same budget as ordinary trajectories, so no assumption about typical records is needed.

Together, the two theorems give a direct separation on the same target family. Choose $\eta=N^{-a}$ with $1/3<a<1/2$. Then $N\eta^2\to\infty$ while $N\eta^3\to0$, so every classical-light count distribution becomes asymptotically maximally distant from the target, whereas the classical sampling error vanishes. Photon counting statistics that cannot be approximated by classical light in this regime can nevertheless be approximated efficiently on a classical computer. The separation is proved on the pair family of Theorem~\ref{thm:lower}. For other interferometers classical light may remain accurate in this regime, and Theorem~\ref{thm:lower} only shows that no classical-light method is accurate on every interferometer, whereas Theorem~\ref{thm:cubic} holds for all of them.

\prlsection{Accuracy at experimental squeezing}
\begin{figure}[t]
 \includegraphics[width=\columnwidth]{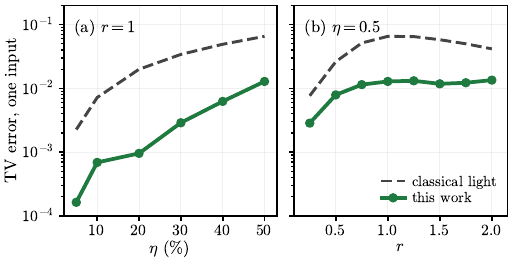}
 \caption{Error for one lossy squeezed input ($N=M=1$). Total variation distance between the exact photon count distribution and the distributions of classical-light replacement by the classical Gaussian state of maximal fidelity~\cite{Qi2020} and of the combined sampler, (a) against the transmission at $r=1$ and (b) against the squeezing at $\eta=0.5$, close to the reported overall efficiency of $0.51$~\cite{Liu2026Jiuzhang}. The reference distance is computed without Monte Carlo sampling. The sampler values are empirical distances from $10^5$ to $2.5\times10^5$ records. The SM gives pointwise $95\%$ statistical upper bounds, including the tail correction.}
 \label{fig:practical}
\end{figure}
For the numerical examples below, we combine the detection approximation with the exact input decomposition of Ref.~\cite{Oh2024}. Each lossy squeezed input is exactly a pure Gaussian core of reduced squeezing followed by classical Gaussian displacement noise. The combined sampler $Q'_{\rm G}$ draws this noise and propagates it through $V$ as a displacement. It then uses the same detection construction on the displaced pure core, matching mean and covariance at each reset. The cost is unchanged, and the update with a mean is given in the SM. Figure~\ref{fig:practical} compares errors on one squeezed input, where the target probabilities can be computed directly. At $r=1$ the empirical error of the combined sampler is below that of the classical reference by a factor between five and twenty across the transmissions. At $\eta=0.5$ the factor is roughly three to five across the squeezing values. Its mean count agrees with the exact value within statistical uncertainty, and random three-mode interferometers give the same ordering. In the examples reported in the SM, $Q_{\rm G}$ is more accurate than the reference for $r\le1$ and less accurate at larger squeezing. The improvement is observed numerically. No uniform analytic error bound is proved for $Q'_{\rm G}$. For the centered reset of Theorem~\ref{thm:cubic} the cubic order is attained by an explicit family of inputs, as shown in the SM.

\prlsection{All input modes squeezed}
When $M=N$, we use an exact amplifier decomposition for the equally squeezed inputs. The isometry is now a unitary $U=V$. Put $W=UU^T$ and
\begin{equation}
 \tau=\frac{\eta m_0}{1+\eta n_0},\quad
 g=\frac{\eta n_0(1-\eta)}{1+\eta n_0},\quad G=1+g.
 \label{eq:amplifier}
\end{equation}
The target is a pure squeezed core of matrix $\tau W$ followed by identical quantum-limited amplifiers of gain $G$~\cite{Ivan2011}. The amplifiers commute with $U$, which makes their count kernels local at the output, and they represent the target rather than hardware added to it. The simplest sampler draws the exact number of core pairs from a negative binomial distribution, $\operatorname{NB}(s,p)$ denoting the distribution with probabilities $\Gamma(s+j)p^s(1-p)^j/[\Gamma(s)\,j!]$ for $j\ge0$ and mean $s(1-p)/p$,
\begin{equation}
 J\sim\operatorname{NB}(N/2,1-\tau^2),\qquad
 \Pr(i,j\mid\text{one pair})=\frac{|W_{ij}|^2}{N},
 \label{eq:pairs}
\end{equation}
assigns the $J$ pair endpoints independently, records their counts $b_i$, adds independent $Y_i\sim\operatorname{NB}(b_i+1,1/G)$, and returns $m_i=b_i+Y_i$. No hafnian, rejection score, or trajectory is needed, and the approximation is solely in the interference between distinct pairs, so the distribution of the total output photon number is exact. With $x=(n_0+1)\eta$,
\begin{equation}
 \begin{gathered}
 \TV(P,Q_{\rm pair})\le\tfrac34\sqrt{Nx^4},\\
 x\le\tfrac1{16},\qquad Nx^4\le\tfrac1{64}.
 \end{gathered}
 \label{eq:pairsbound}
\end{equation}
Amplifier smoothing bounds the omitted cycles, including their phase dependence, as shown in the SM. Equation~\eqref{eq:pairsbound} gives the stronger exponent $3/4$ for the mean surviving photon number, but its error is $O_r(\sqrt N\,\eta^2)$, instead of $O_r(N\eta^4)$. The condition $M=N$ is needed because the same amplifier would create photons in additional vacuum input modes.

\prlsection{Discussion}
Claims of quantum advantage in lossy GBS must rule out simulation within the known efficient regimes. Classical-light replacement~\cite{Qi2020} and the entanglement analysis of Ref.~\cite{Oh2024} give sufficient regimes with the same leading power. Theorem~\ref{thm:lower} establishes a limit for $P$ classical replacement, not for tensor-network methods or classical computation in general. Theorem~\ref{thm:cubic} extends the efficient regime to $N^{2/3}$ surviving photons for every interferometer, and the pair construction extends it to $N^{3/4}$ when every input is equally squeezed. 

Theorem~\ref{thm:cubic} gives a nontrivial asymptotic bound when $\eta\to0$ with $N\eta^3$ bounded. It does not provide a nontrivial uniform bound at fixed nonzero transmission. This is consistent with the hardness arguments for ideal GBS~\cite{Hamilton2017,Grier2022}. Hardness of evaluating noisy output probabilities~\cite{Deshpande2022} concerns a different task from sampling within a fixed total variation error. The latest experiment, with $1{,}024$ squeezed inputs at an overall transmission near $0.5$, has $N\eta^3\approx10^2$, where Theorem~\ref{thm:cubic} gives only the trivial bound of one~\cite{Liu2026Jiuzhang}.

Several questions remain open. Whether typical interferometers also defeat classical light beyond the square-root scale would turn Theorem~\ref{thm:lower} into an average-case statement. Whether $N^{2/3}$, or $N^{3/4}$, is the true limit of polynomial simulation is tied to hardness assumptions, and whether a reset rule with polynomial cost can pass the cubic order of the centered rule is open. Existing approximation algorithms for lossy Fock inputs also give a square-root simulation regime~\cite{Oszmaniec2018,GarciaPatron2019}. The trajectory used here relies on the Gaussian closure of the conditional state, so a detection approximation for lossy Fock states would need a different conditional description, and whether one exists with controlled error is open.

The main theoretical results were developed by the author. GPT-6 and Fable-5.1 were used under the author's direction to assist with exploring and refining mathematical arguments, checking calculations, numerical simulations, and revising the manuscript. The author independently verified the final proofs and takes full responsibility for the results and conclusions. This work was supported by the Institute of Information \& Communications Technology Planning \& Evaluation (IITP) grant funded by the Korean government (Ministry of Science and ICT (MSIT)) (No. RS-2024-00437284, No. RS-2025-02283189), and National Research Foundation of Korea (NRF) grant funded by the Korean government (MSIT) (No. RS-2024-00431768).
\label{maintext-end}
\nocite{Cardin2024,McDiarmid1989,Quesada2020,Quesada2022}
\bibliographystyle{apsrev4-2}
\bibliography{references}
\end{document}


\title{\texorpdfstring{Supplemental Material for\\}{Supplemental Material for }
``Classical Simulation of Lossy Gaussian Boson Sampling beyond Square-Root
Regime''}
\author{Youngrong Lim}
\email{sshaep@gmail.com}
\affiliation{Department of Physics, Chungbuk National University, Cheongju 28644, Korea}
\hypersetup{pdftitle={Supplemental Material for Classical Simulation of Lossy Gaussian Boson Sampling beyond Square-Root Regime}}
\maketitle
\raggedbottom

\noindent Sections~\ref{sec:lower} and~\ref{sec:cubic} give the proofs of
Theorems~1 and~2. Section~\ref{sec:extensions} treats nonuniform inputs and
the limits of the centered reset. Section~\ref{sec:pairs} gives the pair
construction for equally squeezed inputs. The separate variant with
displacement noise and its numerical evaluation are described in
Secs.~\ref{sec:refined} and~\ref{sec:comparison}.

We use vacuum covariance $\id/2$ and write $D_{\rm
tr}(\rho,\sigma)=\tfrac12\|\rho-\sigma\|_1$ and
$\TV(P,Q)=\tfrac12\|P-Q\|_1$. Quadratures are ordered as
$(q_1,\ldots,q_M,p_1,\ldots,p_M)$, with commutator matrix
$\Omega$~\cite{Weedbrook2012}. Except in Sec.~\ref{sec:nonuniform}, the
squeezed inputs have common magnitude $r>0$ and survival probability $\eta$.
The other $M-N$ inputs are vacuum. Put
\[
 n_0=\sinh^2r,\quad m_0=\sinh r\cosh r,\quad
 p_r=n_0+m_0,\quad d_r=m_0-n_0.
\]
Then $p_rd_r=n_0$, $p_r-d_r=2n_0$, and $0<d_r<1/2$. Constants in asymptotic
bounds may depend on the fixed $r$.

\section{Proof of Theorem 1}
\label{sec:lower}
We use the notation $\mathcal C_M$, $Q_\sigma$, and $P_{\eta,V}$ of the
main text. The reference measure may be singular, non-Gaussian, and
correlated across all modes, with no energy bound~\cite{Yadin2018}.

\subsection{Two output modes}
The construction uses independent pairs of squeezed inputs. For $N=2K$,
apply
$U_2=2^{-1/2}\bigl(\begin{smallmatrix}1&i\\1&-i\end{smallmatrix}\bigr)$ to
each input pair. Since
$U_2U_2^T=\bigl(\begin{smallmatrix}0&1\\1&0\end{smallmatrix}\bigr)$, the
target is a product of attenuated two-mode squeezed states. One pair has
generating function
\begin{equation}
 G(z,w)=\frac1{1+n_0[1-(1-\eta+\eta z)(1-\eta+\eta w)]}
       =\frac{D_\eta^{-1}}{1-\beta(z+w)-\kappa zw},\qquad
 \begin{gathered}
 D_\eta=1+n_0(2\eta-\eta^2),\\
 \beta=n_0\eta(1-\eta)/D_\eta,\quad \kappa=n_0\eta^2/D_\eta.
 \end{gathered}
 \label{eq:pairgen}
\end{equation}
Write $A=p_{11}=(2\beta^2+\kappa)/D_\eta$ and
$B=p_{20}+p_{02}=2\beta^2/D_\eta$. Then $A-B=n_0\eta^2/D_\eta^2>0$. For a
coherent state with intensities $I,J$,
$q_{11}-q_{20}-q_{02}=-e^{-I-J}(I-J)^2/2\le0$. Every positive mixture
preserves this inequality between count
probabilities~\cite{Lee1990,Caves1991}.

\subsection{Combining the pairs}

Set $h(1,1)=1$, $h(2,0)=h(0,2)=-1$, and $h=0$ otherwise. For $0<t<1$, let
$w_t=1+th$ and $W_t=\prod_{a=1}^K w_t(X_{2a-1},X_{2a})$. At a fixed coherent
amplitude of all modes, the counts are independent and $\E w_t\le1$ on each
pair. Averaging over the amplitude therefore gives $\E_QW_t\le1$ for every
classical reference, even when its pairs are correlated. Under the target
distribution the pairs are independent. Choosing
\begin{equation}
 t_* =\frac{\sqrt A-\sqrt B}{\sqrt A+\sqrt B},\qquad
 \Delta=\tfrac12(\sqrt A-\sqrt B)^2
 \quad\text{gives}\quad
 \E_p w_{t_*}^{-1}=1-\Delta.
\end{equation}
For $0<\eta<1$, weighted Cauchy--Schwarz now yields
\begin{equation}
 \left(\sum_{\bm n}\sqrt{P(\bm n)Q(\bm n)}\right)^2
 \le\E_PW_{t_*}^{-1}\E_QW_{t_*}\le(1-\Delta)^K,
 \qquad \TV(P,Q)\ge1-(1-\Delta)^{N/4}.
 \label{eq:lowerfinite}
\end{equation}
The endpoint $\eta=1$ follows by continuity. With $u=n_0(1-\eta)$,
\begin{equation}
 \Delta=\frac{\eta^2}{2D_\eta^3}
 \left[\sqrt{u^2+n_0(n_0+1)}-\sqrt2\,u\right]^2
 \ge 4c_r\eta^2,\qquad
 c_r=\frac{[\sqrt{2n_0^2+n_0}-\sqrt2\,n_0]^2}{8(1+n_0)^3}.
 \label{eq:sth1}
\end{equation}
The inequality uses $D_\eta\le1+n_0$ and the decrease of the bracket on
$0\le u\le n_0$. Hence $\TV(P,Q)\ge1-e^{-c_rN\eta^2}$. Appended vacuum
outputs can be marginalized; the remaining marginal of a classical state is
still classical. For odd $N$, ignoring one squeezed input gives the same
statement with $N$ replaced by $N-1$.

For the matching upper bound on this family, use the product of the two
thermal marginals in each pair. Two bivariate distributions with identical
marginals have TV at most twice the sum of their probabilities of both
counts being nonzero. These probabilities are bounded by
$\E_pXY=\eta^2(n_0+2n_0^2)$ and $\E_qXY=\eta^2n_0^2$. Summing over pairs
gives $\TV\le\min\{1,N(n_0+3n_0^2)\eta^2\}$ and establishes the claimed
quadratic scale.

\section{Proof of Theorem 2}
\label{sec:cubic}
We first specify the centered sampler $Q_{\rm G}$. We then bound the error
of each Gaussian replacement and sum it over the complete detection record.
The expected arithmetic cost is given at the end of the section.

\subsection{Conditional evolution and sampling}
\label{sec:algorithm}
Let $O$ be the realification of $V$ and $\mathsf P$ the projector onto the
occupied covariance subspace. Then
\begin{equation}
 O=\begin{pmatrix}\Re V&-\Im V\\\Im V&\Re V\end{pmatrix},\quad
 \mathsf P=OO^T,\quad
 K_0=\eta O\operatorname{diag}(p_r\id_N,-d_r\id_N)O^T.
\end{equation}
The projector $\mathsf P$ has rank $2N$ and commutes with $\Omega$. Let
$\Pi_i$ select the two quadratures of physical output $i$, and
$\ell_i=\frac12\Tr\Pi_i\mathsf P$. Thus $\sum_i\ell_i=N$. All covariances
below are supported perturbations of vacuum, $\Sigma=\id/2+K$, with
$K=\mathsf P K\mathsf P$.

PNR readout can be represented by complete monitored absorption, with one
jump $a_i$ for each photon detected in output $i$. While no click occurs, a
Gaussian state remains Gaussian and obeys
\begin{equation}
 \dot K=-K-K^2,\qquad
 F_t(K)=e^{-t}K[\id+(1-e^{-t})K]^{-1},
 \quad S_K(t)=\det[\id+(1-e^{-t})K]^{-1/2}.
 \label{eq:noclick}
\end{equation}
Here $S_K(t)$ is the probability of no click during the next interval of
length $t$. At a click in $i$,
\begin{equation}
 \lambda_i=\tfrac12\Tr\Pi_iK,\quad
 \Delta_i=\frac{K\Pi_iK}{\lambda_i},\quad K\leftarrow K+\Delta_i.
 \label{eq:sjump}
\end{equation}
The covariance after photon subtraction is exact~\cite{Walschaers2017}.
The Gaussian reset replaces only the higher cumulants by their Gaussian
values, as in Gaussian trajectory approximations~\cite{Verstraelen2018}.
Since $\Delta_i\succeq0$, the new covariance is physical. A detector with
zero rate is never selected.

At each step the record ends with probability
$S_K(\infty)=\det(\id+K)^{-1/2}$. Otherwise draw $\xi$ uniformly from
$(S_K(\infty),1)$ and solve $S_K(t)=\xi$. Evolve $K$ with $F_t$, select the
detector in proportion to its rate at that time, and apply the reset.
Repeating these steps gives the count in each output. No count or covariance
cutoff is imposed.

For numerical implementation, put $u=1-e^{-t}$ and $\widetilde K=e^tK$. Then
$d\widetilde K/du=-\widetilde K^2$ on the finite interval $0\le u<1$.

\subsection{Covariance change after a detection}
\label{sec:jump}
The covariance increment need not be small along every record. We control
it relative to a positive shift of $K$. For $0<\eta<1$, put
\begin{equation}
 b=\frac\eta{1-\eta},\qquad \delta=\eta d_r,\qquad
 Y_t=K_t+b e^{-t}\mathsf P.
 \label{eq:yshift}
\end{equation}

The shift is used only in the proof and does not change the sampler. The flow $F_t$ is operator monotone on the physical domain, and every reset
adds a positive matrix. Hence $K_t\succeq F_t(K_0)$. Put $v=e^{-t}$,
$u=1-v$, and
\begin{equation}
 p(u)=\frac{\eta p_r}{1+u\eta p_r},\qquad
 d(u)=\frac{\eta d_r}{1-u\eta d_r}.
\end{equation}
The eigenvalues of $F_t(K_0)$ are $vp(u)$ and $-vd(u)$ on the two occupied
real eigenspaces. The trace of $\Pi_i$ on either eigenspace is $\ell_i$.
This follows from the realification of $V$ and is the point where the
optical structure is used.

Writing $p_1=p(1)$ and $d_1=d(1)$, direct substitution gives
\begin{equation}
 b=\frac{2p_1d_1}{p_1-d_1}\ge2d_1\ge2d(u).
\end{equation}
Consequently $Y\succeq bv\mathsf P/2$, $K\succeq-\delta\mathsf P$, and
$K^2\preceq Y^2$. The last relation is valid because $K$ and $Y$ commute on
the occupied subspace. Restricting inverses to that subspace,
\begin{equation}
 KY^{-1}K=K-bv\mathsf P+(bv)^2Y^{-1}.
\end{equation}
The baseline resolvent bound implies
\begin{equation}
 \Tr\Pi_iKY^{-1}K\le2\lambda_i+v\ell_i q(u),\quad
 q(u)=-2b+\frac{b^2}{p(u)+b}+\frac{b^2}{b-d(u)}.
\end{equation}
The function $q$ is increasing and $q(1)=0$, so $q(u)\le0$. Dividing by the
rate gives $\Tr(Y^{-1}\Delta_i)\le2$. Since
$Y^{-1/2}\Delta_iY^{-1/2}\succeq0$, its norm is bounded by its trace, giving
$\Delta_i\preceq2Y$. Finally, summing $\lambda_i\Delta_i=K\Pi_iK$ over the detector label
uses $\sum_i\Pi_i=\id$. We have therefore obtained
\begin{equation}
 \Tr(Y^{-1}\Delta_i)\le2,\qquad
 0\preceq\Delta_i\preceq2Y,\qquad
 \sum_i\lambda_i\Delta_i=K^2.
 \label{eq:rels}
\end{equation}
All inverses act on $\operatorname{ran}\mathsf P$. No positive lower bound
on an output occupation is needed, and the estimate does not depend on the
number of vacuum inputs.

\subsection{Error of one Gaussian replacement}
\label{sec:local}
For a centered Gaussian $\rho$ with covariance $\Sigma$, define the
displacement generator by
$\chi_{\mathscr D_\Delta\rho}(z)=-\tfrac12z^T\Delta z\,\chi_\rho(z)$.
Photon subtraction and the Gaussian with the same covariance satisfy
\begin{equation}
 \rho_i^-=(\id+\mathscr D_{\Delta_i})\rho,\qquad
 \rho_{{\rm G},i}=e^{\mathscr D_{\Delta_i}}\rho,\qquad
 e_i:=D_{\rm tr}(\rho_i^-,\rho_{{\rm G},i})
 \le C_*[\Tr(\Sigma^{-1}\Delta_i)]^2.
 \label{eq:localtrace}
\end{equation}
Take a Gaussian purification with covariance
$\Gamma=\bigl(\begin{smallmatrix}\Sigma&C\\C^T&A\end{smallmatrix}\bigr)$,
which satisfies $\Sigma-CA^{-1}C^T=\Omega_s\Sigma^{-1}\Omega_s^T/4$. For a system
displacement $z$, choose the ancillary generator that minimizes the variance
of the combined quadrature $Q_z$. The variance is $z^T\Sigma^{-1}z/4$ and
the system displacement is unchanged. Ancillary commutators vanish after the
partial trace. Taylor's integral remainder and contraction of the trace norm
give
\begin{equation}
 \tfrac12\|[e^{\mathscr D}-\id-\mathscr D]\rho\|_1
 \le\tfrac14\|\mathscr D^2\rho\|_1,\qquad
 \mathscr D_{\rm tot}^{\,2}
 =\tfrac1{12}\E_{z\sim\mathcal N(0,\Delta)}\operatorname{ad}_{Q_z}^{4}.
 \label{eq:disptaylor}
\end{equation}
The second identity follows from Wick's rule for commuting displacement
commutators.

A Gaussian unitary takes a pure Gaussian state to vacuum. For a centered
quadrature of variance $v$, the fourth commutator divided by $v^2$ therefore
reduces to the two blocks
\begin{equation}
 \begin{pmatrix}12&12\sqrt2&2\sqrt6\\12\sqrt2&12&0\\2\sqrt6&0&0\end{pmatrix}
 \oplus
 \begin{pmatrix}-24&-4\sqrt6\\-4\sqrt6&0\end{pmatrix}.
\end{equation}
If $\lambda_-<0$ is the unique negative root of
$\lambda^3-24\lambda^2-168\lambda+288=0$, their total trace norm is
$24+8\sqrt{15}-2\lambda_-$. Using
$\E(z^T\Sigma^{-1}z)^2\le3[\Tr(\Sigma^{-1}\Delta)]^2$ in
Eq.~\eqref{eq:disptaylor} gives
\begin{equation}
 C_* =\frac{24+8\sqrt{15}-2\lambda_-}{256}
 =0.2680684565\ldots<\widehat C_*:=\frac{268069}{10^6}.
 \label{eq:cstar}
\end{equation}
The intermediate derivatives can be defined with finite Fock projections.
The displayed Gaussian moments allow these projections to be removed. The
resulting constant is independent of the number of modes.

To sum the local costs, use the fact that $K,Y,\Sigma$ commute on
$\operatorname{ran}\mathsf P$. Combining Eqs.~\eqref{eq:rels}
and~\eqref{eq:localtrace} gives
\begin{equation}
 \sum_i\lambda_i e_i
 \le 2C_*\Tr(YK^2\Sigma^{-2}).
 \label{eq:sumcost}
\end{equation}
To see this, apply $(\Tr AB)^2\le\Tr(B)\Tr(A^2B)$ with
$B=Y^{-1/2}\Delta_iY^{-1/2}$ and $A=Y\Sigma^{-1}$. Then use $\Tr B\le2$ and
sum over the detector label.

\subsection{Termination and total error}
\label{sec:hybrid}
The energy $\mathcal E_t=\Tr K_t/2$ satisfies $\calG\mathcal E_t=-\mathcal
E_t$. Localization shows that $e^t\mathcal E_t$ is a nonnegative local
martingale and hence a supermartingale. Since $\sum_i\lambda_i=\mathcal
E_t$, the total number $L$ of clicks obeys
\begin{equation}
 \E L\le\mathcal E_0=N\eta n_0,\qquad
 \Pr(L>k)\le\frac{\mathcal E_0}{k},\qquad
 \Pr(\text{a click after }T)\le\mathcal E_0e^{-T}.
 \label{eq:termination}
\end{equation}
Thus the process is nonexplosive and terminates almost surely. The
supermartingale argument proves an upper bound on the mean, not equality.

To compare the output distributions, first fix a finite time and a finite
click cap. Retain one outcome for an unfinished record. Consider hybrids
that follow the candidate for the first $j$ clicks and use exact physical
readout afterward. Two adjacent hybrids share the candidate prefix and the
probabilities of the next click. They differ only at that click, where
$\rho_i^-$ is replaced by $\rho_{{\rm G},i}$.

Every event of the exact suffix is represented by an effect $0\preceq
E\preceq\id$. Its probability changes by at most $|\Tr[E(\rho_i^-
-\rho_{{\rm G},i})]|\le e_i$. Summing over adjacent hybrids weights this
error by the candidate prefix probabilities. The compensator of the counting
process turns the sum over clicks into a time integral.
Equation~\eqref{eq:termination} and monotone convergence remove the two
analytical caps. Thus
\begin{equation}
 \TV(P_{\eta,V},Q_{\rm G})
 \le\E_{Q_{\rm G}}\int_0^\infty\sum_i\lambda_i e_i\,dt.
 \label{eq:hybrid}
\end{equation}
The same comparison holds when the switch to exact readout occurs at a
stopping time of the candidate process. Forgetting the auxiliary click times
cannot increase TV, so the bound applies to the PNR record. The algorithm
imposes neither analytical cap.

\subsection{Cubic error bound}
\label{sec:budget}
The accumulated error is controlled by one positive covariance quantity. Set $s_t=be^{-t}$ and
\begin{equation}
 \Phi_t=\Tr(Y_tK_t^2)=\Tr K_t^3+s_t\Tr K_t^2\ge0.
 \label{eq:polyphi}
\end{equation}
Positivity follows from $Y\succeq0$ and $[Y,K]=0$. A vacuum direction
contributes zero. The exact generator, including $\dot s=-s$, is
\begin{equation}
 \calG\Phi=-3\Phi+\sum_i\lambda_i
 \Tr[(3Y-2s\mathsf P)\Delta_i^2+\Delta_i^3].
 \label{eq:polygen}
\end{equation}
The deterministic contribution $-\Tr[(3K^2+2sK)K^2]$ cancels the part
of the jumps linear in $\Delta_i$ by Eq.~\eqref{eq:rels}. No commutation of
$K$ and an individual jump is required.

For $0\preceq\Delta\preceq2Y$,
\begin{equation}
 \Tr(Y\Delta^2)\le2\Tr(Y^2\Delta),\qquad
 \Tr\Delta^3\le4\Tr(Y^2\Delta).
 \label{eq:polyineq}
\end{equation}
To prove the first relation, write $\Delta=Y^{1/2}BY^{1/2}$, $0\preceq
B\preceq2\id$, in a $Y$ eigenbasis. Symmetrize the terms
$y_i^2y_j|B_{ij}|^2$, use
$y_i^3+y_j^3-y_i^2y_j-y_j^2y_i=(y_i-y_j)^2(y_i+y_j)\ge0$, and then
$B^2\preceq2B$. Trace $\Delta^2$ against $\Delta\preceq2Y$ for the second
relation. Thus
\begin{equation}
 \calG\Phi\le-3\Phi+10\Tr(Y^2K^2)
 \le(-3+10\|Y\|)\Phi,
 \qquad
 \sum_i\lambda_i e_i\le A_{r,\eta}\Phi,
 \quad A_{r,\eta}=\frac{2C_*}{(1/2-\eta d_r)^2}.
 \label{eq:polydrift}
\end{equation}

To include rare records with large covariance changes, choose $b<a<3/10$, set
$\gamma=3-10a$, and let $\mathcal T=\inf\{t:\|Y_t\|\ge a\}$. This stopping
time is used only in the proof. Dynkin's formula for the stopped process and
nonnegativity give
\begin{equation}
 \E[\Phi_{\mathcal T}\mathbf1_{\mathcal T<\infty}]
 +\gamma\E\int_0^{\mathcal T}\Phi_t\,dt\le\Phi_0.
 \label{eq:stopped}
\end{equation}
At a crossing, $Y$ has an eigenvalue $y\ge a$ and $s\le b$, so
$\Phi_{\mathcal T}\ge a(a-b)^2$. Between jumps $Y$ decreases; at a crossing
jump $Y+\Delta\preceq3Y$. This bounds overshoots and justifies stopped
localization. If $\mathcal T=0$, the resulting bound is trivial.

The distance from the target to the stopped hybrid is at most
$A_{r,\eta}\E\int_0^{\mathcal T}\Phi_tdt$. The hybrid and the original
sampler agree until $\mathcal T$. Their distance is therefore bounded by the
crossing probability. Equation~\eqref{eq:stopped} bounds the sum of these
contributions, giving
\begin{equation}
 E_{\rm simple}:=\Phi_0\max\left\{\frac1{a(a-b)^2},
 \frac{A_{r,\eta}}{3-10a}\right\},\qquad
 \TV(P,Q_{\rm G})\le\min\{1,E_{\rm simple}\}.
 \label{eq:simplecert}
\end{equation}
The sampler does not compute the exact suffix used in this comparison.

The initial value is
\begin{equation}
 \Phi_0=N\eta^3\left[8n_0^3+6n_0^2+
 \frac{4n_0^2+2n_0}{1-\eta}\right].
\end{equation}
Taking $a=8/27$, $\eta\le1/10$, $C_*<1/3$, and $1/2-\eta d_r\ge(1-\eta)/2$
yields the explicit choice
\begin{equation}
 C_{3,0}(r)=\frac{19683}{200}\left[8n_0^3+\frac{94}{9}n_0^2+\frac{20}{9}n_0\right]
 \label{eq:c3}
\end{equation}
for $0<\eta\le1/10$. Define $C_3(r)=\max\{C_{3,0}(r),10^3\}$. For
$\eta>1/10$, the right side of the theorem is one. At $\eta=0$ the sampler
returns vacuum exactly.

The fixed threshold $a=8/27$ requires $\eta<8/35$. An admissible $a<3/10$
exists whenever $\eta<3/13$. The smaller interval $\eta\le1/10$ above is
used to obtain the explicit coefficient in Eq.~\eqref{eq:c3}.

\subsection{Arithmetic cost}
\label{sec:work}
Store the realification $O\in\mathbb R^{2M\times2N}$ and the supported
covariance matrix $X\in\mathbb R^{2N\times2N}$, where $K=OXO^T$. For
detector $i$, let $R_i$ contain rows $i$ and $M+i$ of $O$ and put
$A_i=R_i^TR_i$. Each $A_i$ has rank at most two and $\sum_iA_i=\id_{2N}$.
The exact reduced updates are
\begin{equation}
 \lambda_i=\tfrac12\Tr(A_iX),\qquad
 \Delta X_i=XA_iX/\lambda_i,\qquad
 F_t(X)=e^{-t}X[\id+(1-e^{-t})X]^{-1}.
 \label{eq:reducedimplementation}
\end{equation}
Sylvester's determinant identity reduces the no-click determinant to $2N$
dimensions. Hence added vacuum modes do not increase the size of the
dynamical covariance matrix.

A conservative implementation diagonalizes $X$ before each waiting-time draw
at cost $O(N^3)$ and rotates the stored detector rows at cost $O(MN^2)$.
Each survival evaluation then costs $O(N)$, and rates at the chosen time
cost $O(MN)$. The rank-two reset costs $O(N^2)$. Including the terminal
draw, there are $L+1$ steps, with $\E L\le N\eta n_0$. If $b_{\rm inv}$
bounds the number of scalar survival evaluations per inversion, the expected
work is
\begin{equation}
 O\!\left(MN+(1+N\eta n_0)
       [N^3+MN^2+N b_{\rm inv}]\right),
 \qquad \text{memory }O(MN+N^2).
 \label{eq:runtimeexplicit}
\end{equation}
Equation~\eqref{eq:runtimeexplicit} counts scalar inversion as an arithmetic
primitive. It is an operation count, not a bit complexity bound or a runtime
measurement. If a numerical implementation satisfies $\TV(\widehat Q,Q_{\rm
G})\le\epsilon_{\rm num}$, its total error is at most the optical bound plus
$\epsilon_{\rm num}$, including rounding errors.

\section{Extensions and limits of the centered reset}
\label{sec:extensions}
\subsection{Nonuniform squeezing and loss}
\label{sec:nonuniform}
The same scalar shift gives a conservative extension to source dependent
parameters. Let source $k$ have squeezing $r_k$ and survival probability
$\eta_k$, and define
\[
 n_k=\sinh^2r_k,\quad
 b=\max_k\frac{\eta_k}{1-\eta_k},\quad
 \delta=\max_k\eta_kd_{r_k}.
\]
For $0<\eta_{\max}<1$, use $Y_t=K_t+be^{-t}\mathsf P$. The baseline
eigenvalues in Sec.~\ref{sec:jump} become $vp_k(u)$ and $-vd_k(u)$. Each
corresponding real eigendirection has detector weight $|V_{ik}|^2$. The
resolvent estimate becomes
\begin{equation}
 \Tr\Pi_iKY^{-1}K\le2\lambda_i+
 v\sum_k|V_{ik}|^2q_k(u),\qquad
 q_k(u)=-2b+\frac{b^2}{p_k(u)+b}+\frac{b^2}{b-d_k(u)}.
\end{equation}
For each nonvacuum source,
$b_k=2p_k(1)d_k(1)/[p_k(1)-d_k(1)]=\eta_k/(1-\eta_k)$. In addition,
\[
 \partial_bq_k(1)=-\frac{p_k(1)^2}{[b+p_k(1)]^2}
                 -\frac{d_k(1)^2}{[b-d_k(1)]^2}\le0.
\]
Since $b\ge b_k$, $q_k(u)\le q_k(1)\le0$. Zero sources follow by continuity.
Thus Eq.~\eqref{eq:rels} still holds. The initial budget is
\begin{equation}
 \Phi_0=\sum_k\left[\eta_k^3(8n_k^3+6n_k^2)
              +b\eta_k^2(4n_k^2+2n_k)\right].
 \label{eq:nonuniformbudget}
\end{equation}
For $\eta_{\max}\le1/10$, the proof of Sec.~\ref{sec:budget} gives
\begin{equation}
 \TV(P,Q_{\rm G})\le
 \min\left\{1,C_{3,0}(r_{\max})\eta_{\max}\sum_k\eta_k^2\right\}.
 \label{eq:nonuniformbound}
\end{equation}
The same trivial extension uses $C_3(r_{\max})$ outside this interval. The
scalar shift gives $\eta_{\max}\sum_k\eta_k^2$, rather than
$\sum_k\eta_k^3$.

\emph{Additional output loss.} Attenuation of output $i$ with transmission
$t_i$ is implemented after sampling by
$n_i\mapsto\operatorname{Bin}(n_i,t_i)$. This is exactly loss followed by
PNR readout, and the common thinning kernel contracts TV. The bound based on input loss is unchanged; this argument does not sharpen it using the additional
output losses.

\subsection{Cubic error of the centered reset}
\label{sec:whycubic}
The cubic order in Theorem~2 is attained by its centered reset rule.
For $U=\id$, the modes factorize. Write $p_\eta$ for
the exact count law of one mode and $q_\eta$ for the centered sampler. Their
probabilities of three counts obey
\begin{equation}
 p_\eta(3)=A_r\eta^3+O_r(\eta^4),\qquad
 q_\eta(3)=B_r\eta^3+O_r(\eta^4),\qquad
 A_r=n_0^3+\tfrac32 n_0m_0^2.
\end{equation}
At leading order, one reset maps $n\mapsto2n+m^2/n$ and $m\mapsto3m$. Hence
\[
 n_1=2n_0+m_0^2/n_0,\qquad m_1=3m_0,\qquad
 n_2=2n_1+m_1^2/n_1,\qquad B_r=n_0n_1n_2/6.
\]
The factor $1/6$ comes from the three ordered detection times. Direct
substitution yields
\begin{equation}
 B_r-A_r=2n_0^3+2n_0^2+n_0/3>0.
\end{equation}
Coarse graining to the binary event of exactly three counts gives an affinity
deficit $\tfrac12(\sqrt{B_r}-\sqrt{A_r})^2\eta^3+O_r(\eta^4)$. Products over
the $N$ independent modes then imply
\begin{equation}
 \TV(P,Q_{\rm G})\ge1-e^{-\widetilde c_rN\eta^3}
\end{equation}
for sufficiently small $\eta$ and some $\widetilde c_r>0$.

\subsection{Local obstruction for centered Gaussian mixtures}
Let $\rho$ be centered Gaussian with covariance $\Sigma$, and let
$\Delta=\Delta_i$ be the increment of Sec.~\ref{sec:local}. The exact
conditional characteristic function is
\[
 \chi_{\rho_i^-}(z)=\left(1-\tfrac12z^T\Delta z\right)
                    e^{-z^T\Sigma z/2}.
\]
Consider a positive mixture of centered Gaussians
$\sigma=\sum_jp_j\rho_{\Sigma_j}$ with $\sum_jp_j\Sigma_j=\Sigma+\Delta$.
Set $a=z^T\Delta z/2$ and $b=z^T\Sigma z/2$. Convexity gives
$\chi_\sigma(z)\ge e^{-a-b}$. The Weyl operator is unitary, so
\begin{equation}
 D_{\rm tr}(\rho_i^-,\sigma)\ge
 \tfrac12|\chi_{\rho_i^-}(z)-\chi_\sigma(z)|
 \ge\tfrac12e^{-b}(e^{-a}-1+a).
 \label{eq:centeredmixturelower}
\end{equation}
The lower bound is of second order in $\Delta$ and concerns one conditional
state. It does not cover mixtures with displacements or, by itself, imply a
bound on the final PNR error of another adaptive algorithm.

\section{Pair construction when all inputs are equally squeezed}
\label{sec:pairs}
Here $M=N$ and every input has the same squeezing and loss. The common
amplifier representation gives a second sampling method, based on
independently placed pairs rather than Gaussian resets.

\subsection{Amplifier representation}
Set $W=UU^T$, $\alpha=N/2$,
\begin{equation}
 \tau=\frac{\eta m_0}{1+\eta n_0},\quad q=\tau^2,\quad
 g=\frac{\eta n_0(1-\eta)}{1+\eta n_0},\quad G=1+g.
\end{equation}
The exact covariance identities
\begin{equation}
 G\frac{q}{1-q}+g=\eta n_0,\qquad
 G\frac{\tau}{1-q}=\eta m_0
\end{equation}
express the target as a pure Gaussian core of squeezing matrix $\tau W$
followed by identical quantum-limited amplifiers. Equal gains commute with
passive interference. The corresponding output count kernel follows from the amplifier Kraus operators~\cite{Ivan2011}
\begin{equation}
 T_{\bm b}(\bm m)=\prod_i\mathbf1_{m_i\ge b_i}
 \binom{m_i}{b_i}\frac{g^{m_i-b_i}}{G^{m_i+1}}.
 \label{eq:nb}
\end{equation}
For $\eta=0$ the output is vacuum. For $\eta=1$, $g=0$ and the amplifier is
the identity. Assigning a nontrivial amplifier to an additional vacuum input
would change the target.

\subsection{Sampling and total photon number}
Draw $J\sim\operatorname{NB}(\alpha,1-q)$ with
\begin{equation}
 p_j=\Pr(J=j)=(1-q)^\alpha\frac{(\alpha)_j}{j!}q^j.
 \label{eq:pairnumber}
\end{equation}
Choose $J$ independent ordered pairs with probability $|W_{ik}|^2/N$.
Accumulate their endpoints in $\bm b$ and sample $\bm m$ from
Eq.~\eqref{eq:nb}. Denote the output law by $Q_{\rm pair}$. For real shape
$s>0$, the negative binomial convention is $\Pr(j)=(s)_j p^s(1-p)^j/j!$,
with mean $s(1-p)/p$.

Computing $W$ and the row sampling tables costs $O(N^3)$ arithmetic and
$O(N^2)$ storage. Each sample then costs $O(N+J\log N)$, where $\E J=\alpha
q/(1-q)$.

The exact core pair number is Eq.~\eqref{eq:pairnumber}, obtained by summing
$N$ squeezed-vacuum pair counts of shape
$1/2$~\cite{Hamilton2017,Kruse2019}. Conditional on $J$, total amplifier
offspring is $\operatorname{NB}(N+2J,1/G)$ independent of the endpoint
positions. Consequently target and candidate have the same generating function for the total count
\begin{equation}
 (G-gz)^{-N}\left[\frac{1-q}{1-q\{z/(G-gz)\}^2}\right]^\alpha
 =\left[(1+\eta n_0(1-z))^2-\eta^2m_0^2(1-z)^2\right]^{-N/2}.
 \label{eq:totalpgf}
\end{equation}
The distribution of the total count is therefore exact, although the spatial
distribution need not be. For $J=0,1$, the spatial law is also exact. This
gives the finite parameter bound
\begin{equation}
 \TV(P,Q_{\rm pair})\le\Pr(J\ge2)=1-(1-q)^\alpha(1+\alpha q).
 \label{eq:tail1}
\end{equation}

\subsection{Amplifier smoothing}
\label{sec:quartic}
To bound the omitted interference, we compare with a Poisson distribution
of independent pairs. For a diagonal formal variable $Z$, put
$\mathcal Y=WZW^\dagger Z$. This is distinct from the shifted covariance
$Y_t$. The exact core and an auxiliary Poisson pair proposal have generating
functions
\begin{equation}
 P_c=(1-q)^{N/2}\det(\id-q\mathcal Y)^{-1/2},\qquad
 Q_c=\exp[q(\Tr\mathcal Y-N)/2],\qquad
 R:=\log(P_c/Q_c)=\sum_{j\ge2}R_{2j},\quad
 R_{2j}=\frac{q^j}{2j}(\Tr\mathcal Y^j-N).
 \label{eq:cycles}
\end{equation}
The cycles retain the complex phases of $W$~\cite{Cardin2024}. The Poisson
proposal is used only for the error comparison. The sampler itself uses the
exact pair number.

Write $\mathcal A_g$ for Eq.~\eqref{eq:nb}, $\pi=T_{\bm0}$, and $L_{\bm b}$
for the law of $\bm b$ photons with their geometric offspring, so $T_{\bm
b}=L_{\bm b}*\pi$. Let $\mathcal F$ denote the signed coefficient sequence
of $e^R$. The exact amplified law is $h*\mathcal A_g\mathcal F$, where
$h=\E_{\bm b\sim Q_c}L_{\bm b}$. Convolution contracts the signed $\ell_1$
norm.

The amplifier kernel gives an estimate on the resulting count probabilities. For one mode with $g>0$, define
\begin{equation}
 \mu_{b,m}(k)=g^{-k}\frac{\binom m{b+k}}{\binom mb},\qquad
 \zeta^2=G/g.
\end{equation}
Factorial moments of the negative-binomial offspring give
\begin{equation}
 \E_{T_b}\mu_{b,m}(k)\mu_{b,m}(l)
 =\sum_{j=0}^{\min(k,l)}\binom kj\binom lj\frac{j!}{(b+1)_j}\zeta^{2j}.
 \label{eq:gram}
\end{equation}
At $b=0$ this is the circle inner product of $(1+\zeta e^{i\theta})^k$ and
$(1+\zeta e^{i\theta})^l$. Tensorization and Parseval therefore yield, for a
finite signed polynomial $E$,
\begin{equation}
 \|\mathcal A_g E\|_1\le\|E(\bm1+\zeta e^{i\bm\theta})\|_2,
 \label{eq:torus}
\end{equation}
where $\|\cdot\|_2$ denotes the normalized torus norm.

To estimate the cycle terms, set
\begin{equation}
 A=q\zeta^2,\quad y=\zeta^{-2},\quad
 B_d(y)=\sum_{l=0}^d\binom dl^2\frac{y^{d-l}}{l+1},\quad
 \rho=A(1+\sqrt y)^2<1,\quad
 d_* =\frac{A^2(1+\sqrt y)^3}{1-\rho},\quad v=Nd_*^2.
 \label{eq:torusparams}
\end{equation}
For an analytic $f$ on the closed unit polydisc, coefficient orthogonality
gives
\begin{equation}
 \|f-f(0)\|_2^2\le2\int_0^1t\,\E_\theta
 \sum_i|\partial_i f(te^{i\bm\theta})|^2\,dt.
\end{equation}
The right side weights each nonconstant coefficient by
$\sum_i\alpha_i^2/|\bm\alpha|\ge1$. Symmetry and unitarity of $W$ imply
\begin{equation}
 \partial_{z_i}R_{2j}=q^j[\mathcal Y^{j-1}WZW^\dagger]_{ii},\qquad
 \sum_i|\partial_{z_i}R_{2j}|^2\le q^{2j}\Tr|Z|^{4j-2}.
\end{equation}
The second relation uses diagonal-to-Frobenius and Schatten H\"older
inequalities, and integrating the radial expansion gives
\begin{equation}
 \|R_{2j}\|_2^2\le NA^{2j}B_{2j-1}(y),\quad
 \|R-R_4\|_2\le\frac{\sqrt N A^3\sqrt{B_5(y)}}{1-\rho},\quad
 \|R-R_4-R_6\|_2\le\frac{\sqrt N A^4\sqrt{B_7(y)}}{1-\rho}.
 \label{eq:cycletails}
\end{equation}

It remains to control products of the cycle terms. Use shifted variables $F(\bm w)=R(\bm1+\zeta\bm w)$,
so $F(\bm0)=0$, and define
\[
 D_j=F(w_1,\ldots,w_j,0,\ldots,0)
     -F(w_1,\ldots,w_{j-1},0,\ldots,0).
\]
Then $F=\sum_jD_j$, each $D_j$ is analytic in $w_j$, and $D_j|_{w_j=0}=0$.
Integrating $|\partial_{w_j}F|\le d_*$ along the radius gives $|D_j|\le
d_*$. For a subharmonic test function, apply the harmonic majorant first in
the last coordinate, then in the preceding coordinates. This compares $F$
with a sum $Z_*$ of independent uniform circle increments of radius $d_*$.
Rotation symmetry and the multinomial expansion give $\E|Z_*|^{2k}\le
k!v^k$. Applying the comparison to $|e^z-1|^2$ and $|e^z-1-z|^2$ gives
\begin{equation}
 \|e^R-1\|_2\le\sqrt{e^v-1},\qquad
 \|e^R-1-R\|_2\le\sqrt{e^v-1-v}.
 \label{eq:exponential}
\end{equation}
A slightly larger polydisc with $q\max_i|z_i|^2<1$ provides absolute
convergence, so the identities for finite polynomials pass to the infinite
residual.

\subsection{Error bound and surviving photon number}

Isolate $R_4$ before applying Eq.~\eqref{eq:exponential}. The amplified
Poisson pair proposal then has error at most
\begin{equation}
 E_{\rm Pois}=\tfrac12\left[
 \sqrt N A^2\sqrt{B_3(y)}+
 \frac{\sqrt N A^3\sqrt{B_5(y)}}{1-\rho}+\sqrt{e^v-1-v}\right].
 \label{eq:poiscert}
\end{equation}
The exact negative-binomial pair number is the compound Poisson variable
$J=\sum_{k\ge1}kJ_k$, where $J_k\sim\operatorname{Pois}(\alpha q^k/k)$
independently. Coupling it to $J_1$ gives
\begin{equation}
 \delta_{\rm NB}=1-\exp\{\alpha[\log(1-q)+q]\}
 \le\frac{Nq^2}{4(1-q)}.
\end{equation}
The endpoint and amplifier kernels are the same in this comparison.
The full TV distance for $Q_{\rm pair}$ is therefore bounded by the minimum
of one, the right side of Eq.~\eqref{eq:tail1}, and
$\delta_{\rm NB}+E_{\rm Pois}$.

Let $x=(n_0+1)\eta\le1/16$ and $Nx^4\le1/64$. Exact parameter identities
imply
\begin{equation}
 A\le\tfrac{16}{15}x,\quad y\le\tfrac1{16},\quad
 \rho\le\tfrac5{48},\quad d_*\le\tfrac{320}{129}x^2,\quad v<\tfrac1{10}.
\end{equation}
The rational estimates $\sqrt{B_3(1/16)}\le27/40$,
$\sqrt{B_5(1/16)}\le77/100$, and $\frac12\sqrt{e^v-1-v}\le3v/8$ give
\begin{equation}
 E_{\rm Pois}\le\frac{48}{125}\sqrt N x^2+
 \frac{630784}{1209375}\sqrt N x^3+
 \frac{12800}{5547}Nx^4.
\end{equation}
Since $q\le x^2$ and $\sqrt{Nx^4}\le1/8$,
\begin{equation}
 \TV(P,Q_{\rm pair})
 \le\frac{651030344}{884053125}\sqrt{Nx^4}
 \le\tfrac34\sqrt{Nx^4}.
 \label{eq:nbconstant}
\end{equation}
For fixed $r$ and accuracy, take $\eta$ proportional to $N^{-1/4}$. The mean
surviving count is then proportional to $N^{3/4}$. At $\eta=0$, both
distributions are vacuum.

The error is $O_r(\sqrt N\eta^2)$, not $O_r(N\eta^4)$. No rejection score is
required by this construction.

\subsection{Exact sampling of a few pair sectors}
Fix a small $j_*$. For $J=j\le j_*$, repeat row and column $i$ of $W$
exactly $b_i$ times to form $W_{\bm b}$, $|\bm b|=2j$. Exact and
independent-pair conditional laws are
\begin{equation}
 p_j(\bm b)=\frac{j!}{(\alpha)_j}\frac{|\haf W_{\bm b}|^2}{\bm b!},\qquad
 q_j(\bm b)=\frac{j!}{\alpha^j}\frac{\haf(|W_{\bm b}|^{\circ2})}{\bm b!}.
 \label{eq:sector}
\end{equation}
The $M_j=(2j-1)!!$ perfect matchings give $|\haf W_{\bm b}|^2\le
M_j\haf(|W_{\bm b}|^{\circ2})$ by Cauchy--Schwarz. Accept a proposed endpoint vector with probability
\begin{equation}
 a_j(\bm b)=\frac{|\haf W_{\bm b}|^2}{M_j\haf(|W_{\bm b}|^{\circ2})},\qquad
 R_j=(\E_{q_j}a_j)^{-1}=M_j\frac{\alpha^j}{(\alpha)_j}\le M_j.
 \label{eq:sectoraccept}
\end{equation}
A zero denominator has zero proposal probability. Keep $J$ fixed while
retrying the endpoints. Other exact GBS methods can be
used~\cite{Quesada2020,Quesada2022}, but the largest hafnian has size
$2j_*\times2j_*$.

For $J>j_*$ return the independent-pair endpoint draw; do not discard the
sample. Apply Eq.~\eqref{eq:nb} in either case. Then
\begin{align}
 \TV(P,Q^{(j_*)})&\le1-(1-q)^\alpha
               \sum_{j=0}^{j_*}\frac{(\alpha)_j q^j}{j!},\label{eq:sectortail}\\
 \E[\text{endpoint proposals}]&=1+\sum_{j=2}^{j_*}p_j(R_j-1).
 \label{eq:sectorcost}
\end{align}
The entire amplifier tail and the exact total count distribution are
retained. Fixed $j_*$ gives polynomial expected cost. For example, $N=1000$,
$r=1$, $\eta=0.03$, and $j_*=6$ give a tail below $0.00056$ with fewer than
$41$ expected endpoint proposals. This bound from finitely many exact sectors
does not establish the $N^{3/4}$ exponent, because the core pair number
grows with $N$ in that regime.

\section{A variant with displacement noise}
\label{sec:refined}
Combining the exact displacement decomposition of Ref.~\cite{Oh2024} with
Gaussian resets gives the sampler $Q'_{\rm G}$. It retains the sampled
displacement and matches both mean and covariance after each click.
Theorem~2 applies to $Q_{\rm G}$, not to this variant.

\subsection{Input decomposition}
A lossy squeezed input with quadrature variances $V_\pm=\tfrac12(\eta
e^{\pm2r}+1-\eta)$ is the mixture
\begin{equation}
 \rho_{\eta,r}=\int\frac{d\beta}{\sqrt{2\pi\sigma^2}}\,e^{-\beta^2/2\sigma^2}\,D_q(\beta)\,|u\rangle\langle u|\,D_q(\beta)^\dagger,\qquad
 e^{-2u}=2V_-,\qquad \sigma^2=V_+-\tfrac12e^{2u},
 \label{eq:ohdec}
\end{equation}
where $|u\rangle$ is the squeezed vacuum with squeezing $u$ and $D_q(\beta)$
displaces the antisqueezed quadrature by $\beta$. The noise variance is
nonnegative because $V_+V_-\ge1/4$. The core carries the whole deficit of
the squeezed variance, so $u$ is the quantity in the fidelity bound of
Ref.~\cite{Qi2020} and the noise is one dimensional. For $N$ inputs the displacements are
independent, and after the interferometer the state is the pure Gaussian
core with covariance
\begin{equation}
 K_c=O\operatorname{diag}(p_u\id_N,-d_u\id_N)O^T,\qquad p_u=\sinh^2u+\sinh u\cosh u,\qquad d_u=\sinh u\cosh u-\sinh^2u,
\end{equation}
displaced by $\bar x_0=O(\bm\beta,\bm0)^T$. Since $u\approx\eta d_r$ for
small $\eta$, the core occupation of an output is
$n_c\approx\ell_i\eta^2d_r^2$, while its displacement carries
$\sigma^2\ell_i/2\approx\ell_i\eta n_0$ photons.

The refined sampler $Q'_{\rm G}$ draws $\bm\beta$, forms $(\bar x_0,K_c)$,
and runs the monitored absorption of Sec.~\ref{sec:cubic} on a Gaussian
state that now carries a mean. Between clicks the state remains Gaussian.
After a click on output $i$ the conditional state is replaced by the
Gaussian with the same mean and the same covariance.

\subsection{Conditional moments}
Let $\rho$ be Gaussian with mean $\bar x$ and covariance $\Sigma=\id/2+K$. Let $b_i$
annihilate output $i$ and let $\Pi_i$ select its two quadratures. Applying
Wick's theorem to $\langle b_i^\dagger x_kb_i\rangle$ and $\langle
b_i^\dagger x_kx_lb_i\rangle$ gives
\begin{equation}
 \lambda_i=\tfrac12\Tr(\Pi_iK)+\tfrac12\bar x^T\Pi_i\bar x,\qquad
 \bar x^-=\bar x+\frac{v_i}{\lambda_i},\qquad
 \Sigma^-=\Sigma+\frac{K\Pi_iK}{\lambda_i}-\frac{v_iv_i^T}{\lambda_i^2},\qquad v_i=K\Pi_i\bar x .
 \label{eq:dispupdate}
\end{equation}
At $\bar x=0$ this is the update rule of the main text, and a coherent state
($K=0$) is left unchanged. The no-click evolution of duration $t$ with
$h=1-e^{-t}$ is
\begin{equation}
 K_t=e^{-t}K(\id+hK)^{-1},\qquad
 \bar x_t=e^{-t/2}(\id+hK)^{-1}\bar x,\qquad
 S(t)=\det(\id+hK)^{-1/2}\exp\!\big[-\tfrac h2\,\bar x^T(\id+hK)^{-1}\bar x\big],
 \label{eq:dispnoclick}
\end{equation}
Here $S(t)$ decreases from one to the vacuum probability. The mean remains
in $\operatorname{ran}O$, so store $K=OXO^T$ and $\bar x=O\bar w$. The
expected arithmetic work has the same polynomial scaling as
Eq.~\eqref{eq:runtimeexplicit}. The energy $\mathcal E=\Tr K/2+\|\bar
x\|^2/2$ has generator $-\mathcal E$, proving an upper bound on the expected
count, not equality. The negative term $-v_iv_i^T/\lambda_i^2$ in
Eq.~\eqref{eq:dispupdate} prevents direct use of the relative jump bound and
cubic budget of Sec.~\ref{sec:cubic}.

\subsection{Small displacements}
A small pure core does not make every reset small. If the displacement
intensity $|\alpha_i|^2$ is comparable to or smaller than the core
occupation $n_{c,i}$, the reset can have order one error. At zero
displacement it replaces $S(u)\ket1$ by a squeezed thermal state, at trace
distance $3/4$. For one mode, the case of Fig.~2 of the main text, the noise
of Eq.~\eqref{eq:ohdec} is one dimensional. In the initial displacement
ensemble this region has probability of order $\sqrt\eta$, and the detection
rate there is of order $\eta^2$. These initial rates suggest a contribution of order $\eta^{5/2}$ but do not
bound the full trajectory error. The numerical improvement in Fig.~2 does
not sharpen the constant in Theorem~2. A core with squeezing $u<s<\tfrac12\log(2V_+)$ and noise in both quadratures with variances $V_+-\tfrac12e^{2s}$ and $V_--\tfrac12e^{-2s}$ may remove this
restriction to one noise quadrature, but no error analysis of that variant is given
here.

\section{Numerical evaluation}
\label{sec:comparison}
\subsection{Probability evaluation}
\label{sec:numericalmethod}
Figure~2 of the main text concerns $N=M=1$. The numerical calculations use a
Fock basis with cutoff $80$ for the target and classical Gaussian reference.
Both samplers use the covariance representation. Each point uses
$2.5\times10^5$ independent records, except $1.5\times10^5$ at $r=1.75$ and
$10^5$ at $r=2$ for $\eta=0.5$.

Define the count bins as $0,1,\ldots,79$ and a separate bin for $n\ge80$.
Let $P^c,Q^c$ be the exact target and candidate laws on these bins, and let
$\widehat Q^c$ be the empirical histogram. Write $\widehat
d=\TV(P^c,\widehat Q^c)$. Folding the tail is not the same as truncating and
renormalizing a state. The statistical bounds below treat $P^c$ as exact; a
numerical error in its evaluation must be added separately.

\subsection{Statistical uncertainty}
For the binned laws, convexity gives $\E\widehat d\ge\TV(P^c,Q^c)$.
Replacing one independent record changes $\widehat d$ by at most $1/n_{\rm
rec}$. McDiarmid's inequality~\cite{McDiarmid1989} therefore yields the
pointwise statement
\begin{equation}
 \TV(P^c,Q^c)\le\widehat d+
 \sqrt{\frac{\log(1/\delta)}{2n_{\rm rec}}}
 \quad\text{with probability at least }1-\delta.
 \label{eq:mcbinned}
\end{equation}
Combining the tail into one bin gives
\begin{equation}
 0\le\TV(P,Q)-\TV(P^c,Q^c)
 \le\min\{P(n\ge80),Q(n\ge80)\}\le P(n\ge80).
 \label{eq:tailfolding}
\end{equation}
Thus the full count distributions satisfy, pointwise at confidence $95\%$,
\begin{equation}
 \TV(P,Q)\le\widehat d+
 \sqrt{\frac{\log20}{2n_{\rm rec}}}+P(n\ge80).
 \label{eq:mcbound}
\end{equation}
The corrections are below $0.0025$, $0.0032$, and $0.0046$ for the three
sample sizes above. At $r=2$ and $\eta=0.5$, the target tail is below
$7\times10^{-4}$, and it is below $2\times10^{-5}$ at the other points.
The confidence bounds apply to each point separately. The improvement
factors quoted in the main text use the empirical distances, not these
confidence bounds.

If $P^c$ is evaluated approximately as $\widetilde P^c$, add $\TV(\widetilde
P^c,P^c)$ to Eq.~\eqref{eq:mcbound}, together with any error in the target
tail bound. A Fock cutoff alone does not bound this numerical error.

\subsection{One and three mode examples}
At $\eta=0.5$, $Q_{\rm G}$ becomes
less accurate than the classical reference above $r\approx1.1$ in these
tests. Over the squeezing values tested, the empirical error of $Q'_{\rm G}$ is
between about $0.003$ and $0.013$. Its sampled mean agrees with the exact
mean within statistical uncertainty. This agreement does not establish
exact mean preservation.

For random interferometers on three modes at $\eta=0.3$, the empirical errors
of the reference, $Q_{\rm G}$, $Q'_{\rm G}$, and $Q_{\rm pair}$ are
$(0.058,0.058,0.014,0.015)$ at $r=1$, and $(0.062,0.136,0.04,0.080)$ at
$r=1.5$. For $Q'_{\rm G}$, both empirical distances are comparable to sampling
noise and do not resolve the true error. These tests on a few modes do not
establish a uniform error bound or compare runtime with tensor network
methods~\cite{Oh2024,Liu2026}.

\subsection{Checks of the formulas and implementation}
\label{sec:validation}
The reduced covariance representation, summed jump identity, cubic generator,
and local cost inequality were checked in exact rational arithmetic for
$18$ Gaussian prefixes and $54$ detector jumps. The complex rational
isometries have $(N,M)=(1,2),(2,3),(2,4)$, and $42$ jumps are noncommuting.
In $18$ single-mode Fock calculations, increasing the dimension from $40$
to $56$ changes the local trace distance by at most $1.4\times10^{-16}$.
The largest ratio to Eq.~\eqref{eq:localtrace} is $0.867$.

For the displaced variant, Eqs.~\eqref{eq:dispupdate} and
\eqref{eq:dispnoclick} agree with the Fock calculations to $10^{-16}$ in one
mode and to approximately $10^{-5}$ at cutoff $11$ in two modes. A trajectory with nonuniform parameters $r_k\in\{0.29,0.46,1.08\}$ and
$\eta_k\in\{0.39,0.15,0.32\}$ reaches $\max_i\Tr(Y^{-1}\Delta_i)=2.0000$
without leaving $Y\succ0$.

Tests of the complete sampler $Q_{\rm G}$ use one or two squeezed inputs in up to
three modes, $r\in\{0.3,0.5,1\}$, and $\eta\in\{0.1,0.2,0.3\}$. With $10^5$
records, the empirical distances range from the sampling resolution,
about $9\times10^{-5}$, to $2.6\times10^{-2}$. Where the bound in
Eq.~\eqref{eq:simplecert} is below one, it exceeds the empirical distance
by more than an order of magnitude.

For one mode at $r=1$ and $\eta=0.3$, a run with $10^6$ records has mean
$0.3795\pm0.0010$, compared with the exact $0.4143$. It gives probability
$1.3\times10^{-4}$ for more than $16$ clicks, compared with $2\times10^{-6}$
for the target. The smaller sampled mean is consistent with Eq.~\eqref{eq:termination},
which gives only an upper bound. These numerical values do not establish
failure of uniform integrability. At $r=0.3$, numerical quadrature gives
$Q_{\rm G}(1)-P(1)\approx0.53n_0\eta^3$ and $Q_{\rm
G}(2)-P(2)\approx-1.05n_0\eta^3$ as $\eta\to0$, consistent with the order
proved for this reset in Sec.~\ref{sec:whycubic}.

These calculations check the formulas at finite Fock dimension and numerical
precision. They do not rigorously bound truncation or rounding errors.
The analytical bounds are proved independently of the numerical tests.

\bibliographystyle{apsrev4-2}
\bibliography{references}